\documentclass[conference]{IEEEtran}

\usepackage{graphicx}
\usepackage{cite}
\usepackage[cmex10]{amsmath}
\usepackage{threeparttable}
\usepackage{amsmath}
\usepackage{amssymb}
\usepackage{diagbox}

\usepackage{graphicx}
\usepackage{epstopdf}

\usepackage{algorithm}
\usepackage{algpseudocode}
\usepackage{multirow}

\def\BibTeX{{\rm B\kern-.05em{\sc i\kern-.025em b}\kern-.08em
    T\kern-.1667em\lower.7ex\hbox{E}\kern-.125emX}}
\begin{document}
\title{An Asymmetric Adaptive SCL Decoder Hardware for Ultra-Low-Error-Rate Polar Codes}
\author{\IEEEauthorblockN{Jiajie Tong, Huazi Zhang, Lingchen Huang, Xiaocheng Liu, Jun Wang}
\IEEEauthorblockA{Huawei Technologies Co. Ltd.\\
Email: \{tongjiajie,zhanghuazi,huanglingchen, liuxiaocheng, justin.wangjun\}@huawei.com}}
\maketitle
\begin{abstract}
In theory, Polar codes do not exhibit an error floor under successive-cancellation (SC) decoding.
In practice, frame error rate (FER) down to $10^{-12}$ has not been reported with a real SC list (SCL) decoder hardware.
This paper presents an asymmetric adaptive SCL (A2SCL) decoder, implemented in real hardware, for high-throughput and ultra-reliable communications.
We propose to concatenate multiple SC decoders with an SCL decoder, in which the numbers of SC/SCL decoders are balanced with respect to their area and latency.
In addition, a novel unequal-quantization technique is adopted.
The two optimizations are crucial for improving SCL throughput within limited chip area.
As an application, we build a link-level FPGA emulation platform to measure ultra-low FERs of 3GPP NR Polar codes (with parity-check and CRC bits).
It is flexible to support all list sizes up to $8$, code lengths up to $1024$ and arbitrary code rates.
With the proposed hardware, decoding speed is 7000 times faster than a CPU core.
For the first time, FER as low as $10^{-12}$ is measured and quantization effect is analyzed.
\end{abstract}

\begin{IEEEkeywords}
Polar codes, A2SCL, Emulation platform, FER.
\end{IEEEkeywords}

\IEEEpeerreviewmaketitle

\section{Introduction}
Polar codes, proposed by Arikan\cite{Arikan}, has been selected by the 5G standards. Polar codes with successive-cancellation (SC) decoding theoretically achieve channel capacity in the asymptotic sense. To improve error-correction performance at short or moderate lengths, SC list (SCL) decoding is proposed by keeping $L$ codeword candidates. Concatenated with cyclic redundancy check (CRC)\cite{CA_polar} or parity check (PC)\cite{PC-polar} bits, the error-correction performance can be further improved.

One advantage of polar codes is that it does not exhibit an error floor when decoded by the SC and SCL algorithms.
This makes Polar codes suitable for applications with stringent error performance requirements.
For some industrial and medical applications, FER is required to be smaller than $10^{-10}$.
However, an efficient hardware solution designed for this purpose has not been reported yet.

It is not easy to achieve this goal in an efficient way, because both decoding latency and throughput should be highly optimized within limited chip area.
To our best knowledge, an ultra-low FER below $10^{-10}$ has not been reported from a real hardware.
Although many efforts have been made to optimize the decoder hardware of Polar codes \cite{parallel_Architecture,6920050,7001058,7337462,7114328,goodbit,fast_and_flexible,Polar_ASIC},
the lowest FER reported in a real hardware is $\approx 10^{-6}$ (not fulfilling the $<10^{-10}$ requirement).
An FPGA emulation platform is designed for ultra-reliable communications \cite{URLLC}, but does not present any hardware-measured FER results.

\subsection{Motivation and Contribution}
To achieve ultra-reliable and high-throughput decoding, we adopt the adaptive SCL decoder framework in \cite{Aaptive_SCL}.
To further improve throughput, we propose an asymmetric adaptive SCL (A2SCL) decoder, based on the observation that SC and SCL decoders exhibit huge differences in terms of the area \& latency, as well as quantization precision. A2SCL mainly adopts the following two techniques:
\begin{enumerate}
  \item Asymmetric deployment: the number of SC and SCL decoders are no longer 1:1 as in the original design, but carefully chosen to reflect their significant difference in terms of area and latency.
  \item Asymmetric quantization: The different demands for data precision between SC and SCL decoders are also exploited to pack as many SC decoders for parallel decoding, yet without FER loss.
\end{enumerate}

In addition, we provide a reference design through an efficient emulation platform in an FPGA, and evaluate the ultra-low FER performance of Polar codes to demonstrate its practical value.
The proposed A2SCL decoder not only achieves FER $\approx 10^{-12}$, but also supports list sizes $1,2,4,8$ with maximum code length $N_{max}= 2^{10}$. The emulation platform has the following features:
\begin{itemize}
\item Integrity: All modules in the link-level emulation such as source vector generator, encoder, modulator, AWGN channel and decoder are executed in the FPGA. The server is only responsible for the code lengths/rate configuration and results collection. 
\item Efficiency: The emulation platform dramatically improves evaluation speed. One FPGA board can be up to 7000 times faster than a CPU core. 
\item Flexibility: The emulation platform supports CA-Polar (up to 24 CRC bits), PC-Polar \cite{PC-polar} (as specified by 3GPP), various rate-matching schemes, list sizes, code lengths and code rates. All these can be configured by the server on the fly.
\item Scalability: A server can manage one or more FPGAs to speed up the emulation. Servers can also form a cluster to further speed up the emulation.
\end{itemize}

With the emulation platform, ultra-low FER performance of Polar codes is measured and the error-correction performance of 3GPP NR Polar codes is evaluated.

\section{Polar Codes}
An $(N,K)$ polar codes has $N$ coded bits and $K$ information bits. The code rate is $R=K/N$. The information bits are assigned to the $K$ most reliable sub-channels, and frozen bits, typically zeros, are assigned to the remaining ones.
The encoding of Polar code is $c=uF ^ {\otimes n}$, where $u$ is the information vector (including information and frozen bits),
$F^{\otimes n}=\bigl[ \begin{smallmatrix} 1 & 0 \\ 1 & 1 \end{smallmatrix} \bigr]^{\otimes n}$
is the transformation matrix, where $^\otimes$ denotes Kronecker power, and $n=\log_2N$.
\subsection{SC-based Decoders}
\begin{figure}
\centering
\includegraphics[width=3.0in]{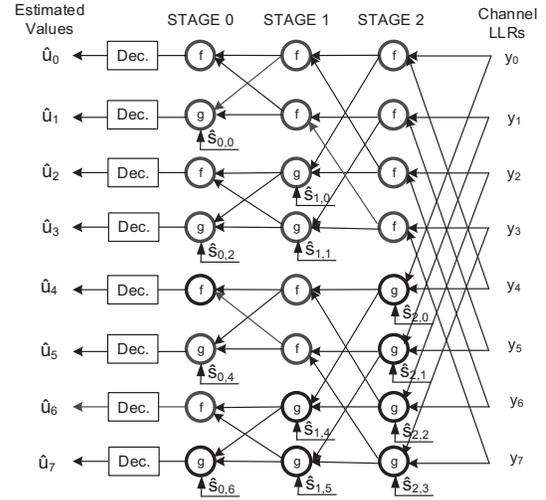} %
\caption{SC decoding graph.}
\label{decoding_graph}
\end{figure}

The decoding graph of SC decoder is shown in Fig. \ref{decoding_graph}. The soft bits propagate from right to left and the hard bits propagate from left to right. The information vector $u$ is decoded sequentially from top to bottom. A hardware-friendly version of soft value updating is carried out in log-likelihood ratio (LLR) domain \cite{7114328}. Two incoming LLRs ($L_{in1}$ and $L_{in2}$) are combined to produce $L_{out}$ with the following f-function
\begin{equation}
L_{out}=sign(L_{in1} \cdot L_{in2}) \cdot min(|L_{in1}| , |L_{in2}|) .
\end{equation}
or g-function
\begin{equation}
L_{out}=L_{in1} + (-1)^{\hat{s}} \cdot L_{in2} ,
\end{equation}
where $\hat{s}$ is the modulo-2 sum of previously decoded bits and is called partial sum (PS).

For an SCL decoder, the decoding process is similar to SC decoder except that it keeps $L$ paths. When making hard decision for each bit, $L$ paths split into $2L$ paths, and the ones with smallest path metric (PM) are kept. For the $l_{th}$ path and bit $u_i$, the LLR of stage $0$ is denoted by $L_{0,i}^l$ and its hard decision is denoted by $\beta_{i}^l$.
The PMs update according to
\begin{equation}
PM_i^l=  \left\{ \begin{array}{lll}
         PM_{i-1}^l,  &  \mbox{if} \   u_i^l = \beta_{i}^l \\
         PM_{i-1}^l+|L_{0,i}^l|,  &  \mbox{otherwise}
         \end{array}\right.
\end{equation}
After all bits are decoded, the path with the smallest PM is selected as the decoding output.

For CRC aided SCL (CA-SCL), the most reliable path that passes CRC check is selected as the decoding output. For parity-check SCL (PC-SCL), each parity bit is decided by its parity function rather than by its LLR. A PC-CA-SCL decoder combines the features of both, if both CRC bits and PC bits are employed. Throughout this work, we implement CA-SCL and PC-CA-SCL decoders.

\section{Asymmetric Adaptive SCL (A2SCL) Decoder}
The original adaptive SCL decoder \cite{Aaptive_SCL} progressively increases the list size until a packet is successively decoded or a maximum list size $L_{\max}$ is reached. Our implementation is built upon a simplified version of \cite{Aaptive_SCL} that has only two decoders, i.e., an SC and an SCL with a given list size. The algorithm is described in Algorithm~\ref{alg:adaptiveSCL}.
\begin{algorithm}
\caption{Simplified Adaptive SCL Decoder:}
\begin{algorithmic}
\State {\bf(1)} Try to decode the incoming packet using SC.
\State {\bf(2)} If the decoded data passes CRC check, go to (4), else go to (3).
\State {\bf(3)} Try to decode the incoming packet using SCL with a fixed list size $L$.
\State {\bf(4)} Compare with the original data and update error counter accordingly. Over.
\end{algorithmic}
\label{alg:adaptiveSCL}
\end{algorithm}

\begin{table}[t]
    \caption{Comparison of area and latency between an SC and SCL decoder with $L=8$}
    \label{SC_vs_SCL8}
    \centering
    \begin{tabular}{|c|c|c|c|c|c|c|}\hline
\multirow{2}{*}{Decoder} & \multicolumn{2}{c|}{Area / Quantization} & \multicolumn{4}{c|}{Latency / Code Rate} \\ \cline{2-7}
                         & 6 bit              & 8 bit              & 1/8      & 1/4     & 1/2      & 3/4      \\ \hline
SC                       & 1.00               & 1.27              & 1.00      & 1.43     & 2.04      & 2.34      \\ \hline
SCL with $L=8$           & 5.06               & 6.32              & 4.69      & 7.39     & 9.94     & 12.1     \\ \hline
    \end{tabular}
\end{table}

Although the software implementation of Algorithm~\ref{alg:adaptiveSCL} is rather straightforward, its hardware implementation is different. One has to take into account the huge difference between an SC decoder and an SCL decoder in terms of hardware resource, as well as their work load at a target signal-to-noise ratio (SNR).

The chip area \& decoding latency comparison between an SC and an SCL decoder ($L=8$) is shown in Table~\ref{SC_vs_SCL8}. The (normalized) measurements are based on our reference ASIC implementations in \cite{Polar_ASIC}, with both SC and SCL decoders optimized to their best efficiency (see details in \cite{Polar_ASIC}). According to the measurements, both the area and latency of an SCL decoder ($L=8$) is up to 6 times of an SC decoder with the same quantization and code rate. If we implement many SCL decoders with different list sizes, both the area efficiency and time efficiency will be very low.

\begin{figure}
\centering
\includegraphics[width=3.15in]{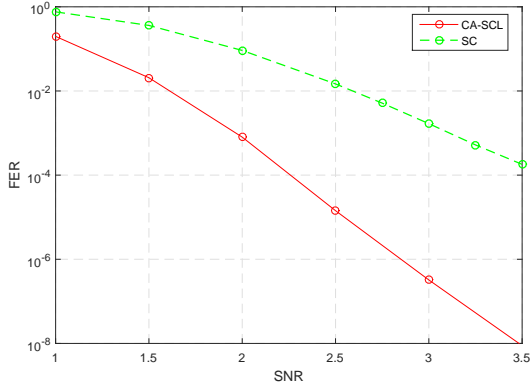} %
\caption{SC vs CA-SCL ($L=8$ with 24 CRC bits)}
\label{sc_vs_scl}
\end{figure}

The work load comparison between an SC and an SCL decoder is given through a case study of ($N=1024, K=512$, 24 CRC bits) Polar codes. As shown in Fig.~\ref{sc_vs_scl}, the required SNR for CA-SCL with $L=8$ to achieve ultra-reliable communications (FER$\leq 10^{-8}$) is around 3.5 dB. In such a high SNR region, an SC decoder already exhibits very small FER ($\sim 10^{-4}$), i.e., only loses one or two packets in 10,000. That means, while SC needs to process all packets, only a small fraction of packets need to be processed by the SCL decoder. This is a huge difference in terms of work load.

Considering the above, a direct implementation of \cite{Aaptive_SCL} would incur very low hardware utilization efficiency. To address this, we propose an asymmetric adaptive SCL (A2SCL) decoder to overcome the above mentioned issues.

\subsection{Asymmetric deployment}
To increase throughput, an A2SCL decoder deploys as many SC decoders as possible. To improve efficiency, A2SCL implements only one SCL decoder (e.g., $L_{\max}=8$)\footnote{In our FPGA platform, we implement a flexible SCL decoder that can be configured to support $L_{\max}=2,4,8$.}, instead of many SCL decoders with different list sizes (e.g., $L=2,4,8$).

\begin{figure}[!htb]
\centering
\includegraphics[width=3.15in]{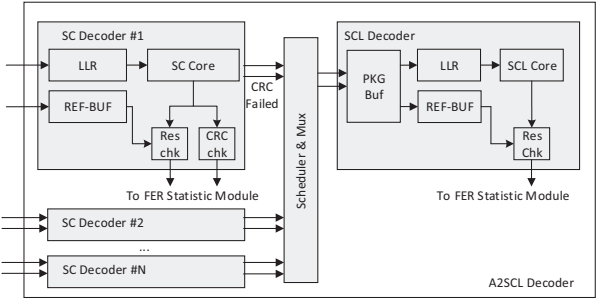} %
\caption{A2SCL decoder hardware architecture}
\label{fig_adapter_SCL_decoder}
\end{figure}

A scheduler with a MUX is used to collect the CRC-failed packets from the SC decoders, and send them to the SCL decoder. Fig. \ref{fig_adapter_SCL_decoder} shows the hardware architecture of the A2SCL decoder. We refer to the different number of SC decoders and SCL decoder as ``asymmetric deployment''.

The SC decoder Core and SCL decoder Core have the similar architecture as described in \cite{Polar_ASIC}, which summarizes some state-of-the-art optimizations over SC and SCL decoders. Both decoders only store intermediate LLRs for every two neighboring stages in the trellis graph shown in Fig.~\ref{decoding_graph}. The ``double-packet mode'' and ``decoded-bit recovery'' features \cite{Polar_ASIC} are enabled to reduce the number of LUT/BRAM/FF modules. The hardware-friendly ``syndrome-check'' \cite{hard_decision_ref} and ``decision-aided'' \cite{goodbit} approaches are adopted to increase the throughput of SC and SCL decoders, respectively.

Assume the work target is FER$<10^{-9}$, in almost all cases, SC decoder's FER$<10^{-3}$ under the target SNR. According to simulation results and real hardware test results\cite{Polar_ASIC}, the SC decoder and SCL decoder's throughput ratio is 5:1. Thus, the SCL decoder can process the failed packets of 200 SC decoders at FER$<10^{-9}$.

The LLR buffer size of the SCL decoder should be larger than those of SC decoders, in case that many SC decoders generate failed packets at the same time.
The following formula evaluates the probability that, during one SCL decoding, the SC decoders have failed $e$ packets.

\begin{equation}\label{prob}
P(e)={\binom {c}{e}} \times FER^e \times (1-FER)^{c - e},
\end{equation}
where $c = \frac{N_{SC}\times 2 \times T_{SCL}}{T_{SC}}$ is the total number of packets processed by the SC decoders during one SCL decoding, $N_{SC}$ is the number of SC decoders in the A2SCL, $T_{SCL}$ and $T_{SC}$ are the decoding time of SCL and SC decoders, respectively.

In our final design, $N_{SC}=18$ SC decoders are implemented in the A2SCL decoder. As mentioned above, $T_{SCL}/T_{SC} = 5$. Assume the SC decoders work at $FER \approx 10^{-3}$, Table~\ref{table_prob} shows the probabilities when the number of SC-failed packets $e$ increases from 0 to 4. According to the table, the probability that $e<3$ is $99.9\%$. Thus, we set the LLR buffer size of the SCL decoder to be 2048 (two packets at maximum), while larger sizes are also allowed.

\begin{table}
  \renewcommand{\arraystretch}{1.15}
    \caption{SC decoding failure probability}
    \label{table_prob}
    \centering
    \begin{tabular}{|c|c|c|c|c|c|}
    \hline
       $e$ failures     &0    & 1  & 2    & 3      & 4   \\
    \hline
      Probability    &83.52\%     & 15.05\%       &1.35\%        &0.08\%      & 0.0035\%  \\
       \hline
    \end{tabular}
    \end{table}

\subsection{Asymmetric quantization}
In a real hardware design, all LLRs are quantized. But the number of quantization bits should be carefully chosen.
More quantization bits improves decoding performance, but requires the extra hardware resource.
Therefore, we choose the fewest number of bits that incurs negligible performance loss.
Moreover, since SCL decoder and SC decoder take different roles in the A2SCL, we propose to employ different quantization widths for them.
The scheme is called asymmetric quantization.

\begin{figure}
\centering
\includegraphics[width=3.15in]{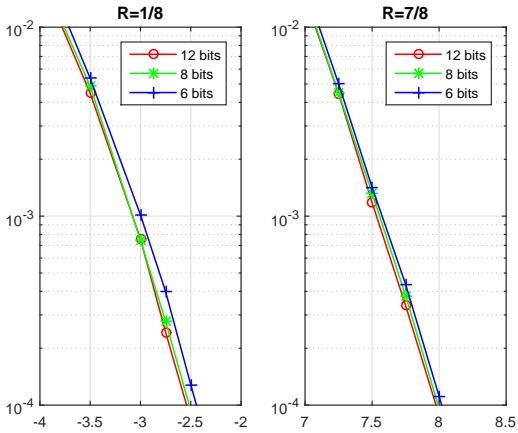} %
\caption{SC FER@Different Quantization Bits}
\label{bit_fer}
\end{figure}

First, an SC decoder should be as fast as possible. For an A2SCL decoder, its SC decoding performance can be relaxed to some extent, because the SCL decoder will take care of the failed packets.
Typically, longer codes require more quantization bits than short ones.
As shown in Fig.~\ref{bit_fer}, the FER curves of $N=1024, K=[1/8,7/8]$ Polar codes with quantization bits $=[6,8,12]$ are almost the same under SC decoding. Accordingly, 6-bits or 8-bits quantization is sufficient for SC decoders.

\begin{figure}
\centering
\includegraphics[width=3.15in]{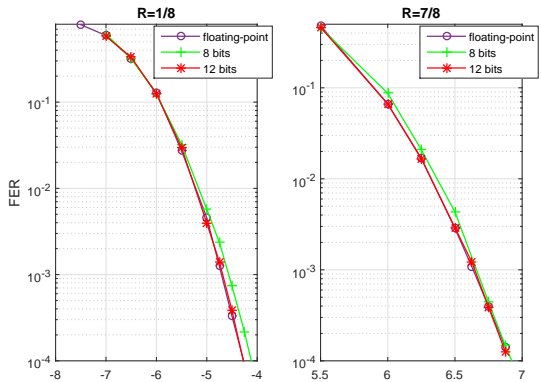} %
\caption{SCL ($L=8$) FER @ Different Quantization Bits}
\label{bit_scl_fer}
\end{figure}

Second, the SCL decoder should yield almost the same performance as a floating-point decoder. We plot the FER curves of $N=1024, K=[1/8,7/8]$ Polar codes under SCL decoding ($L=8$) as reference to show the influence of different quantization bits. As shown in Fig.~\ref{bit_scl_fer}, 8-bits quantization incurs 0.1db loss at maximum, and 12-bits quantization yields the same performance as a floating-point decoder. Accordingly, we adopt 12-bits quantization.

\section{Emulation platform}
\begin{figure}
\centering
\includegraphics[width=0.5\textwidth]{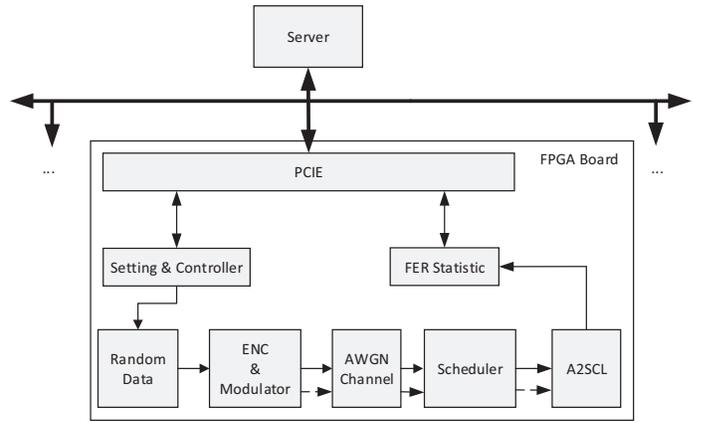}
\caption{The architecture of the A2SCL emulation platform.}
\label{top architecture}
\end{figure}

An overview of our platform is shown in Fig.~\ref{top architecture}. A server can manage one or more FPGA boards via the PCI-E slots. When multiple FPGA boards (constrained by the number of PCI-E slots) are employed, the decoding throughput can be further increased.

A Xilinx xc7vx690t is integrated in the FPGA board. The server is the controller of the platform. The code construction (information sub-channel positions) can be configured by the server to evaluate different code constructions. In addition, code and channel parameters are also configured at the server. According to these configurations, frames are generated, encoded, passed through the AWGN channel and decoded in the A2SCL decoder. The number of decoded frames and frame errors are counted in the FPGA and collected by the server. Finally, the FER curve is displayed on the server.

\subsection{Encoder}
\begin{figure}
\centering
\includegraphics[width=3.15in]{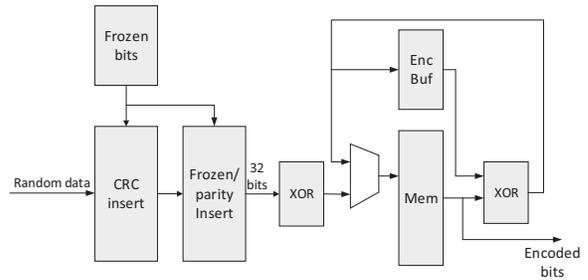} %
\caption{The encoder architecture}
\label{encoder}
\end{figure}

A random bit stream of length $K$ is generated. As shown in Fig. \ref{encoder}, the frozen bits are set to zero, the CRC and parity check bits are inserted.
The pre-coded bits are then fed into a polar encoder.

To achieve high-throughput encoding, every 32 bits are processed in parallel. Specifically, a polar code of length $N$ is split into $N/32$ short codes with length 32. The encoder consists of two parts. At first, a short Polar code of length 32 is encoded and stored into memory. Then, $N/32$ short codes are encoded iteratively with the memory, buffer and the XOR logic.
Intermediate results are stored in the buffer. The size of the buffer is half of $N_{\max}$. At most three frames can be stored in the memory. Therefore, frames can be encoded with this pipelined fashion.
\subsection{AWGN channel}
The AWGN noise sequence is generated by converting a uniform distributed sequence in the range [0,1] using the inverse cumulative distribution function (ICDF) \cite{ICDF}. A 32-bit hardware random number generator is designed based on a 43-bits linear feedback shift register (LFSR), and a 37-bits cellular automata shift register (CASR) \cite{random_data}. The cycle length of the combined generator is close to $2^{80}$.

To reduce the mapping table between white uniform noise and white Gaussian noise, we employ $128$ line segments to approximate the ICDF. Taking advantage of the symmetry of the ICDF, the mapping table is reduced to 64 starting points, terminal points and slopes. In addition, one multiplier and one adder are required to rebuild the ICDF.
 \begin{figure}
\centering
\includegraphics[width=3.15in]{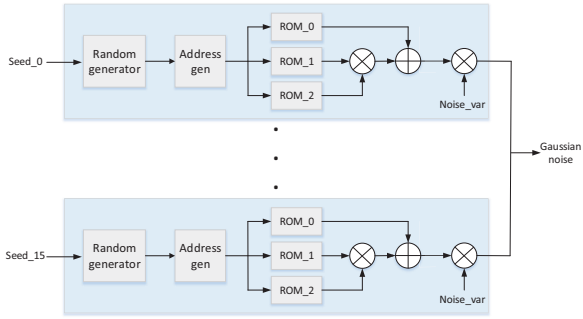} %
\caption{The Gaussian noise generator architecture}
\label{gaussian_gen}
\end{figure}

As shown in Fig. \ref{gaussian_gen}, $16$ AWGN generators with different seeds are combined to provide high throughput. The noise is quantized to $16$ bits.
Experiment results in Section~\ref{sec:performance} also show that the quantized noise has negligible effect on FER performance.

\subsection{Run time balancing}
In the link-level emulation, different modules have different run time to process one packet. Balancing the run time among different modules will benefit the overall operating efficiency. Table \ref{table_cycle} shows the running cycles required by each module in the link-level emulation platform \footnote{SC/SCL cycles are half of the sum run time for two packets due to pipelining.}$^,$\footnote{SC employs syndrome-check acceleration \cite{hard_decision_ref}, the number of cycles is measured @ $FER_{SC} \approx 10^{-3}$; } for different $(N,K)$ case.

Obviously, the running cycles of encoder and AWGN depend on the code length $N$, and SC/SCL decoders depend on both $N$ and $K$. According to the number of running cycles, we integrate the same number of encoder and AWGN modules, and set the ratio of encoders and SC decoders to be $1:2$.
\begin{table}
  \renewcommand{\arraystretch}{1.15}
    \caption{Number of running cycles for different modules}
    \label{table_cycle}
    \centering
    \begin{tabular}{|c|c|c|c|c|}
    \hline
       $(N,k)$          & Encoder  & AWGN    & SC Decoder     & SCL Decoder   \\
    \hline
      (1024,512)         & 97       &76        &221      &1073   \\
       \hline
       (1024,128)           & 97      &76      &108        & 506  \\
       \hline
      (512,256)            & 41      &44         &105          &498  \\
    \hline
      (256,128)           & 21      &28         &66            &261 \\
    \hline
    \end{tabular}
\end{table}

Thanks to the asymmetric architecture, we can integrate more modules within the limited FPGA resource. Our FPGA chip integrates 9 encoders, 9 AWGN channel modules, one A2SCL decoder which include 18 8-bits-quantized SC decoders and one 12-bits-quantized SCL decoder. The resource utilization of each module is shown in Table~\ref{table_source} \footnote{Encoder resource includes the 9 encoder modules. AWGN resource includes the 9 AWGN modules. A2SCL decoder resource includes the 18 8-bit-quantization-SC decoder cores, one 12-bit-quantization-SCL decoder core and the schduler/mux units that connect them.}.
\begin{table}
  \renewcommand{\arraystretch}{1.15}
    \caption{FPGA resource utilization}
    \label{table_source}
    \centering
    \begin{tabular}{|c|c|c|c|c|}
    \hline
                 & Encoder    & AWGN        & A2SCL & Total\\
    \hline
      LUTs         &32991       &98170         &215252    &346413\\
       \hline
       FFs           &20683      &32821           &48420    &101924 \\
       \hline
      RAM            &54         &216            &680.5     &1050.5\\
    \hline
      DSP           &0         &720               &0       &720\\
    \hline
    \end{tabular}
    \end{table}
\subsection{Hardware vs software implementations}\label{sec:performance}
To justify the A2SCL hardware platform, its simulation speed and FER performance are compared with a software counterpart.
The hardware platform utilizes only one FPGA board.
The software implementation is written by C language, and runs on a server that contains 4 Intel Xeon(R) E5-4627 v2@3.30GHz CPUs with 12 cores and 256 GB RAM.
For fairness, the data type of the LLR in the software decoder is short.

\subsubsection{Simulation speed}
\begin{figure}
\centering
\includegraphics[width=3.35in]{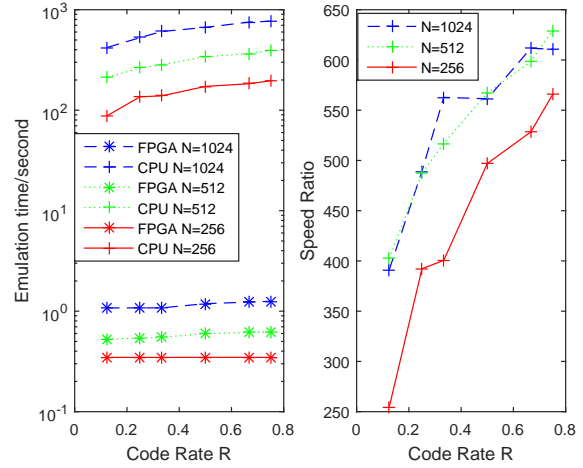} %
\caption{The emulation time of the FPGA platform vs software implementation.}
\label{runtime}
\end{figure}

\begin{figure}
\centering
\includegraphics[width=3.35in]{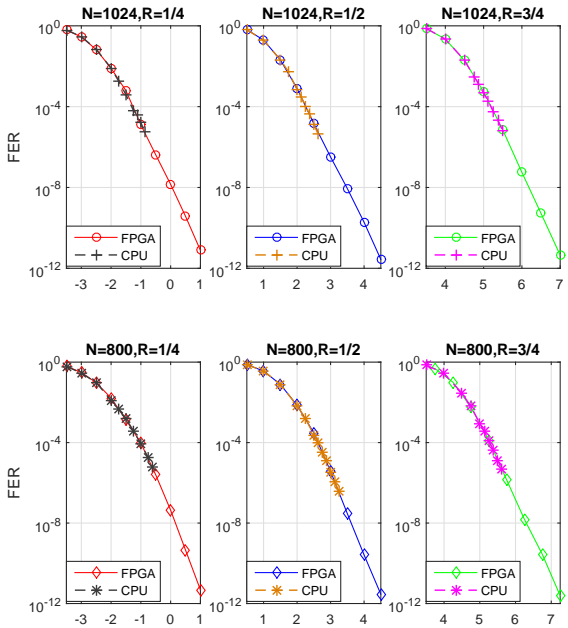} %
\caption{The measured FER performance.}
\label{bler}
\end{figure}

The emulation time of $10^7$ frames by the FPGA platform and software implementations are plotted and compared in the Fig.~\ref{runtime}. We evaluated code lengths $N=[256,512,1024]$, and rates $R=[1/8,1/4,1/3,1/2,2/3,3/4]$. As seen, the emulation time of the FPGA platform is much shorter than the software implementation on CPUs. When $N=1024$ and $R=3/4$, the emulation time of the FPGA platform is about 1.275 seconds. However, the CPUs requires 767.9 seconds. 

Define the speed ratio (SR) as the emulation time of 12 CPU cores divided by that of one FPGA board, also plotted in Fig.~\ref{runtime}.
The highest SR is 611, which means that one FPGA board is 611 times faster than 12 CPU cores.
Converted to one CPU core, a FPGA board is 7332 times faster.
As shown, the emulation platform can greatly reduce emulation time.

\subsubsection{FER performance}
Based on 5G Polar codes with lengths $N=1024$ and $N=800$, we compare the the floating-point results from software and fixed-point results from the FPGA platform.
Due to the very time-consuming floating-point simulation, software results for $FER > 10^{-6}$ are measured. As shown in Fig.~\ref{bler}, the FER results of the FPGA platform perfectly match the floating-point results under various code lengths and code rates.
Note that no error floor is observed from the FPGA platform even when FER results are below $10^{-6}$.
\section{Performance of 5G Polar Codes in 5G {\upshape e}MBB}
Thanks to the FPGA platform, we can now quickly evaluate error-correction performance of 5G Polar codes at FER below $10^{-11}$.

\begin{figure}
\centering
\includegraphics[width=3.05in]{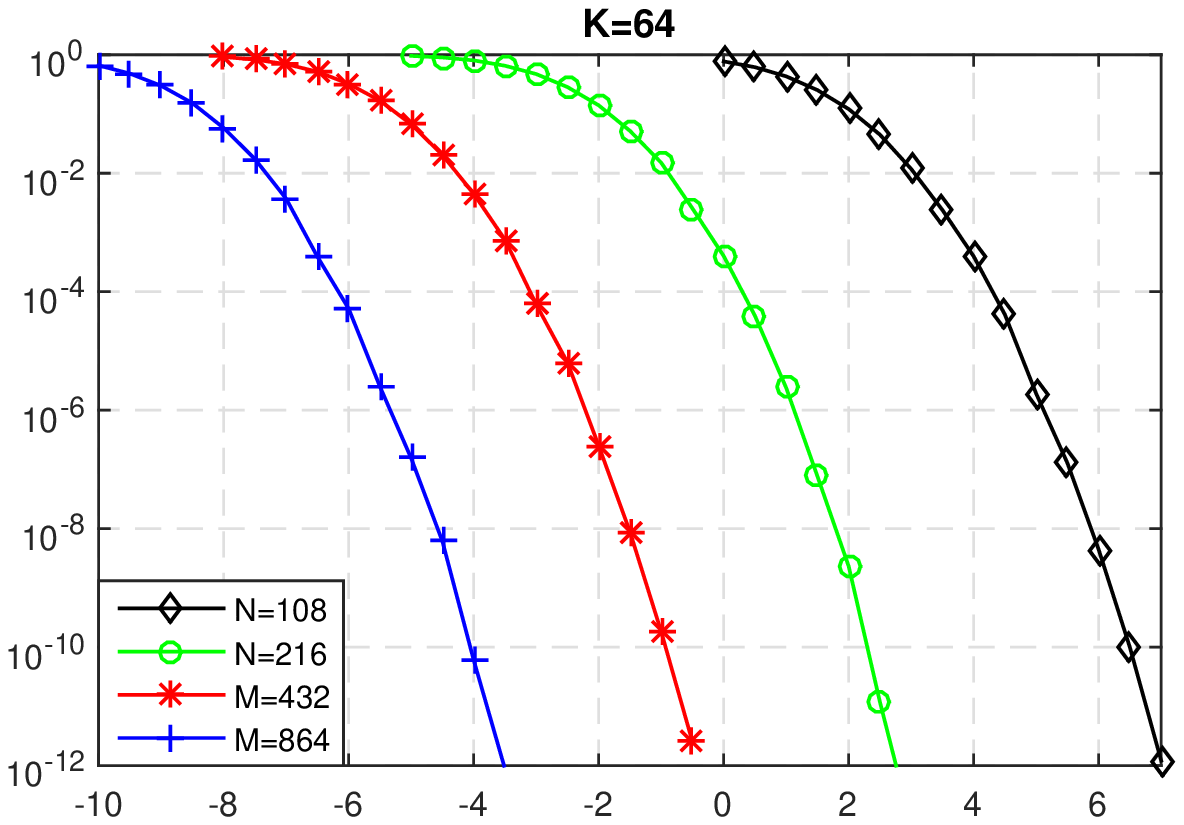} %
\caption{The FER Performance@$k=64$,PDCCH Aggregation Level$=[1,2,4,8]$}
\label{DL64_bler}
\end{figure}

\begin{figure}
\centering
\includegraphics[width=3.05in]{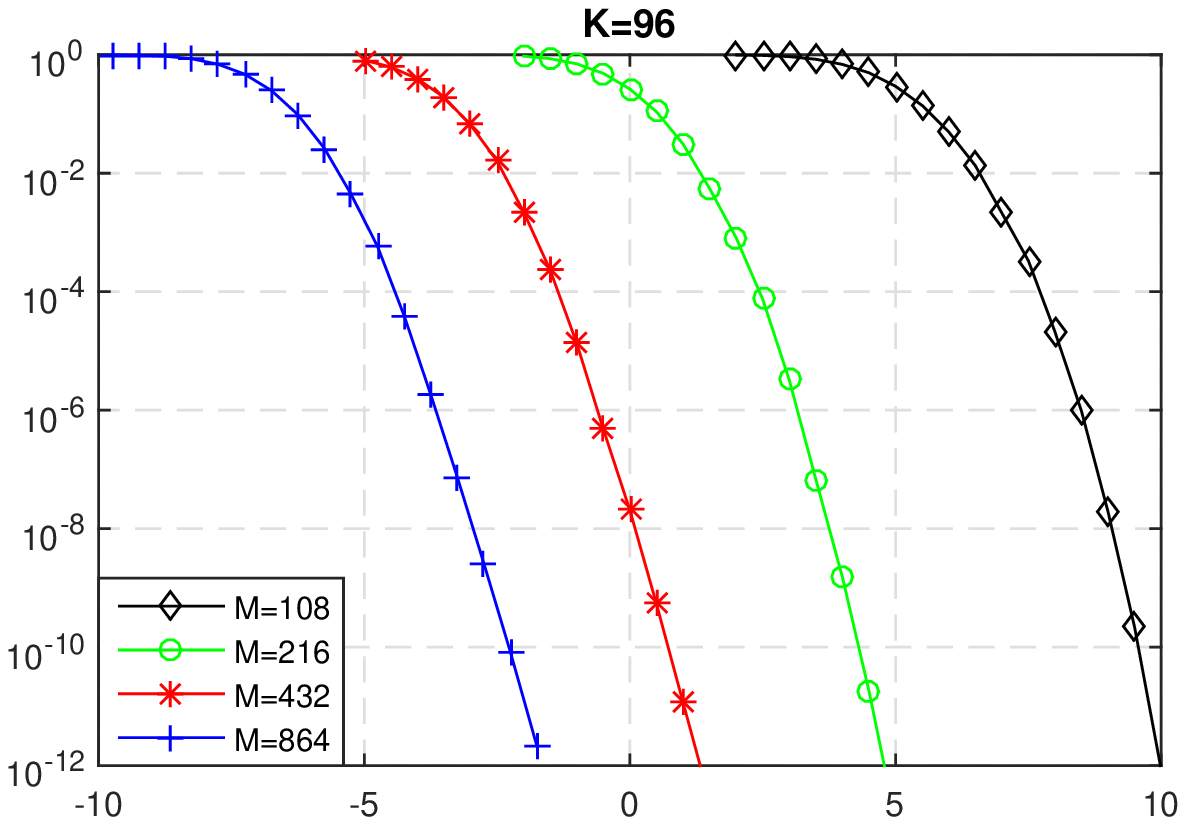} %
\caption{The FER Performance@$k=96$,PDCCH Aggregation Level$=[1,2,4,8]$}
\label{DL96_bler}
\end{figure}

\begin{figure}
\centering
\includegraphics[width=3.05in]{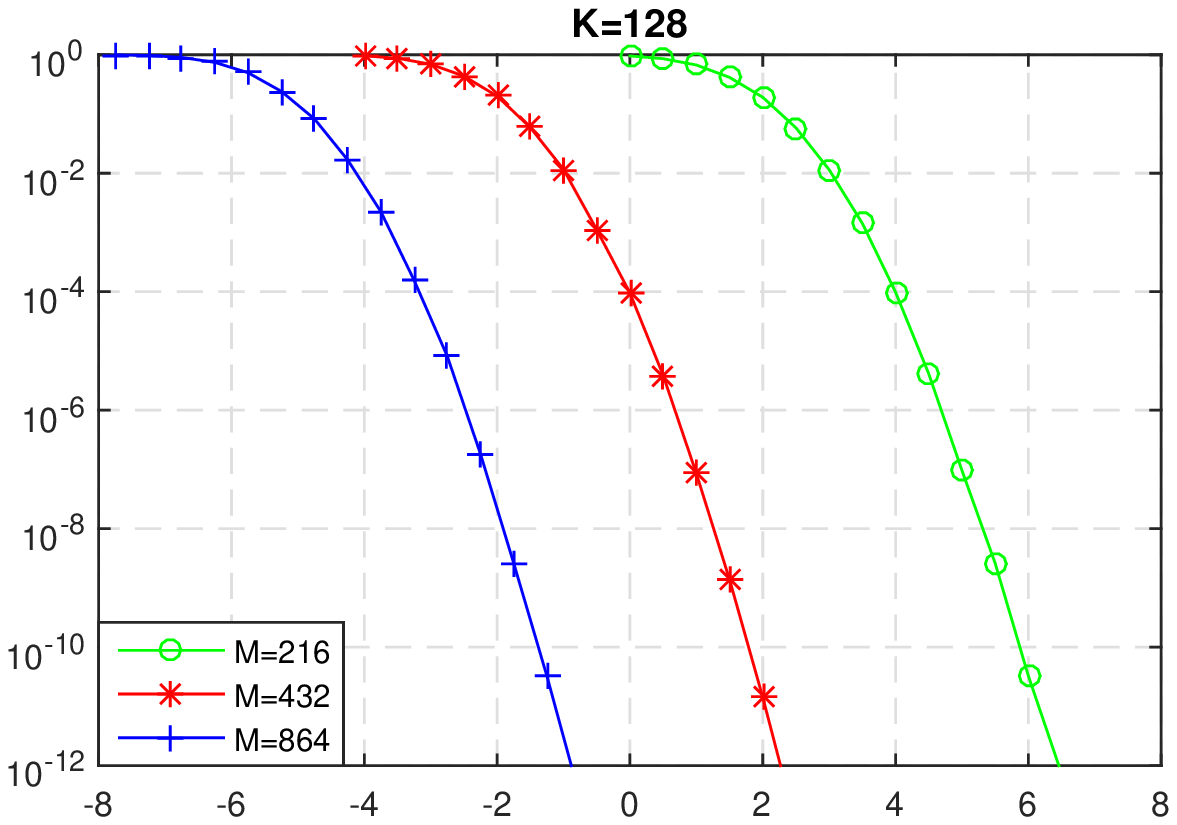} %
\caption{The FER Performance@$k=128$,PDCCH Aggregation Level$=[2,4,8]$}
\label{DL128_bler}
\end{figure}

\begin{figure}
\centering
\includegraphics[width=3.05in]{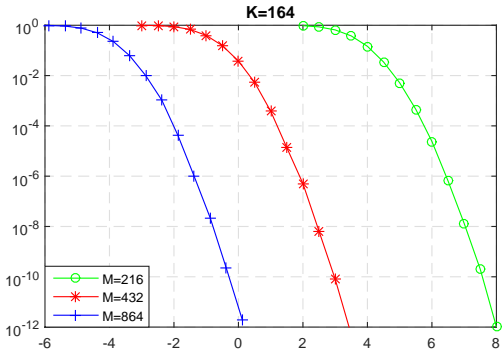} %
\caption{The FER Performance@$k=164$,PDCCH Aggregation Level$=[2,4,8]$}
\label{DL164_bler}
\end{figure}

The typical cases of downlink control information (DCI) are evaluated. For $K=64$, PDCCH aggregation levels $[1,2,4,8]$\footnote{For $K=64$, aggregation level 1, the rate matching is shortening; for aggregation level $[2,4]$, the rate matching is puncturing; for aggregation level 8, the rate matching is repetition.}, the measured FER results are shown in Fig.\ref{DL64_bler}. And we also measured $K=[96,128,164]$, PDCCH aggregation levels $[1,2,4,8]$ the results are shown in the Fig.\ref{DL96_bler}, Fig.\ref{DL128_bler} and Fig.\ref{DL164_bler}.

\section{Conclusion}
In this paper, we present an asymmetric adaptive SCL decoder in real hardware. Equipped with asymmetric deployment and asymmetric quantization, the decoder can provide much higher decoding throughput in a resource-limited FPGA/AISC chip. The A2SCL algorithm, along with all the required link-level modules, is implemented in an FPGA platform. The platform is efficient, flexible and scalable. The emulation speed of one FPGA board is 611 times as fast as 12 CPU cores; converted to one CPU core, a FPGA board is 7332 times faster. Ultra-low FER performance as low as $10^{-12}$ is measured for 5G Polar codes for the first time in real hardware.

\end{document}